\documentclass[aps,prl,twocolumn,superscriptaddress,showpacs]{revtex4-1}
\usepackage[T1]{fontenc}
\usepackage{lmodern}
\usepackage[latin1]{inputenc}
\usepackage[swedish,english]{babel}
\usepackage{amsmath}
\usepackage{amsfonts}
\usepackage{graphicx} 
\usepackage{xfrac}
\usepackage{color}
\usepackage{multirow}
\usepackage{hyperref}

\hypersetup{
	colorlinks = true,
	pdftitle = {},
	pdfauthor = {},
	pdfkeywords = {},
	linkcolor = blue,
	citecolor = blue,
	filecolor = black,
	urlcolor = magenta
}

\newcommand{\dd}{\mathrm{d}}
\newcommand{\ee}{\mathrm{e}}
\newcommand{\img}{\mathrm{i}}

\graphicspath{{./figures/}}

\begin{document}

\title{Elastic scattering of electron vortex beams in magnetic matter}
 
\author{Alexander Edstr\"om}
\affiliation{Department of Physics and Astronomy, Uppsala University, Box 516, 75121 Uppsala, Sweden}

\author{Axel Lubk}
\affiliation{Triebenberg Laboratory, Technische Universit\"at Dresden, Germany}

\author{J\'{a}n Rusz}
\affiliation{Department of Physics and Astronomy, Uppsala University, Box 516, 75121 Uppsala, Sweden}

\date{\today}

\begin{abstract}
Elastic scattering of electron vortex beams on magnetic materials leads to a weak magnetic contrast due to Zeeman interaction of orbital angular momentum of the beam with magnetic fields in the sample. The magnetic signal manifests itself as a redistribution of intensity in diffraction patterns due to a change of sign of the orbital angular moment. While in the atomic resolution regime the magnetic signal is most likely under the detection limits of present transmission electron microscopes, for electron probes with high orbital angular momenta, and correspondingly larger spatial extent, its detection is predicted to be feasible.
\end{abstract}

\maketitle

Rapid developments in nanoengineering call for characterization methods capable to reach high spatial resolution. In this domain, the scanning transmission electron microscope (STEM) provides a broad scale of measurement techniques ranging from Z-contrast \cite{zcontrast} or electron energy-loss elemental mapping \cite{eelsmapping}, differential phase contrast (DPC) \cite{diffpc, Muller2014}, via local electronic structure studies of single atoms \cite{quentin} to counting individual atoms in nanoparticles \cite{countatoms}. As a specific case of high-spatial resolution electron energy loss spectroscopy, an electron magnetic circular dichroism (EMCD) method has been introduced \cite{nature} as an analogue to x-ray magnetic circular dichroism, which is a well established quantitative method of measuring spin and orbital magnetic moments in an element-selective manner \cite{thole,carra}.

Recenly, the introduction of electron vortex beams (EVB) \cite{Uchida2010,Verbeeck2010,McMorran2011}, i.e., beams with nonzero orbital angular momentum, aimed at probing EMCD at atomic spatial resolution. It was shown theoretically that EVBs need to be of atomic size in order to be efficient for magnetic studies \cite{Rusz2013,schatt,emcdc34}. Several methods of generating atomic size electron vortex beams have been proposed \cite{Blackburn2014,monopole,hermes,Pohl2015}, yet an experimental demonstration of atomic resolution EMCD has not been presented in the literature.

An alternative route to utilizing EVBs for magnetic measurements is based on Zeeman interaction between their angular momentum and the magnetic field in the sample. The Pauli equation for an electron with energy $E$ in an electrostatic potential $V(\mathbf{r})$ and a constant magnetic field $\mathbf{B}_\text{c}$ reads
\begin{equation}
	\left[ \frac{\hat{\mathbf{p}}^2}{2m} + \frac{e}{m}(\hat{\mathbf{L}} + 2\hat{\mathbf{S}})\cdot \mathbf{B}_\text{c} -eV(\mathbf{r}) \right] \Psi(\mathbf{r}) = E \Psi (\mathbf{r}), 	
\label{PeqConsB}
\end{equation}
where $-e$ is the electron charge, $m$ is the electron mass, $\hat{\mathbf{p}}=-i\hbar\nabla$ is the momentum operator, $\hat{\mathbf{L}}$ and $\hat{\mathbf{S}}$ are the orbital and spin angular momentum operators, and $\Psi(\mathbf{r})$ is a two-component spinor wavefunction. The second term on the left hand side of Eq.~\ref{PeqConsB} manifests a coupling between the magnetic field and the orbital and spin angular momenta of the electron beam. A previous study has indicated that the effect of spin on elastic scattering is very weak \cite{Rother2009}. Moreover, generating intense spin polarized electron beams remains a technological challenge \cite{tanaka} and so far magnetic field mapping with spin-polarized electrons in the TEM could not be demonstrated. While the spin angular momentum of electrons in the propagation direction is at most $\frac{\hbar}{2}$, EVBs can be generated with very high orbital angular momenta (OAM) \cite{McMorran2011,Saitoh2012,Grillo2015}, which permits an increase of the Zeeman interaction by more than two orders of magnitude.

In this Letter, we show that there is a magnetic contrast in elastic scattering of EVBs originating from the enhanced Zeeman interaction of the beam OAM with magnetic fields in the sample. The described effect is sensitive to magnetic fields parallel to beam-direction, which would complement holographic or DPC methods measuring the in-plane components of the magnetic field.

For a realistic description of magnetism in a solid, taking into account merely a constant magnetic field is insufficient. Hence, we consider a stationary Pauli equation with a non-uniform magnetic field \cite{Strange} $\mathbf{B}(\mathbf{r})=\nabla\times\mathbf{A}(\mathbf{r})$ and corresponding vector potential $\mathbf{A}(\mathbf{r})$ in Coulomb gauge, $\nabla \cdot \mathbf{A}(\mathbf{r}) = 0$. Due to large acceleration voltages commonly applied in TEM, a relativistically corrected electron mass $m=\gamma m_0$ is used. Subsequently, as in the derivation of the conventional multislice method \cite{Cowley1957}, we introduce a paraxial approximation \cite{kirkland} via the substitution
\begin{equation}
	\Psi(\mathbf{r}) =  \ee^{\mathrm{i} k z} \begin{pmatrix} \psi_\uparrow(\mathbf{r}) \\ \psi_{\downarrow}(\mathbf{r}) \end{pmatrix},
\end{equation}
and neglect the second derivatives of the envelope functions $\psi_{\uparrow,\downarrow} (\mathbf{r})$ with respect to the beam propagation direction $\mathbf{k}=(0,0,k)$. The resulting two-component paraxial Pauli equation reads \cite{Asquared}
\begin{widetext}\begin{equation}
	\frac{\partial}{\partial z} \begin{pmatrix} \psi_\uparrow(\mathbf{r}) \\ \psi_{\downarrow}(\mathbf{r}) \end{pmatrix} = \frac{i m}{\hbar}\left( \hbar k + e A_z \right)^{-1} \left\{ \frac{\hbar^2}{2m}\nabla_{xy}^2   + \frac{ie\hbar}{m}\mathbf{A}_{xy}\cdot\nabla_{xy} - \frac{\hbar k e A_z}{m} - \frac{e}{m}\hat{\mathbf{S}}\cdot\mathbf{B} + eV \right\}  \begin{pmatrix} \psi_\uparrow(\mathbf{r}) \\ \psi_{\downarrow}(\mathbf{r}) \end{pmatrix} \equiv \hat{H} \begin{pmatrix} \psi_\uparrow(\mathbf{r}) \\ \psi_{\downarrow}(\mathbf{r}) \end{pmatrix},
\label{paraxialPeq}
\end{equation}\end{widetext}
which upon setting $\mathbf{A} = \mathbf{B} = 0$ reduces to the paraxial Schr\"odinger equation \cite{kirkland} for each of the spin components $\psi_{\uparrow,\downarrow}$ separately. Eq.~\ref{paraxialPeq}, however, represents a system of two differential equations coupled via an interaction of the spin of the probe with the magnetic field in the sample. It can be integrated slice-by-slice according to \cite{Cai2009,Cowley1957}
\begin{equation}
\begin{split}
\mathbf{\psi}(x,y,z+\Delta z) & = \hat{Z}\{ \ee^{\int_z^{z+\Delta z} \hat{H}(x,y,z') \dd z' } \}  \mathbf{\psi}(\mathbf{r}) \approx \\
 & \approx \sum_{n=1}^\infty \frac{\Delta z^n}{n!}{\hat{H}}^n(\mathbf{r}) \mathbf{\psi}(\mathbf{r}),
\end{split}
\label{multislicesol}
\end{equation}
where $\hat{Z}$ is Dyson's $z$-ordering operator. Similar computational methods were recently discussed for the fully relativistic case of the Dirac equation \cite{Rother2009} and in the context of spin polarization devices \cite{Grillo2013}. 

In the following numerical simulations we constructed the electrostatic potential from tabulated values of independent atoms \cite{kirkland}, whereas the magnetic vector potential and the corresponding magnetic field are obtained from density function theory (DFT) in the following way.

In a crystal, $\mathbf{B}$ consists of a constant part due to the saturation magnetization $\mathbf{B}_\text{c} = \mu_0 \mathbf{M}_\text{s}$ and a periodic part $\mathbf{B}_\text{p}$ that averages to zero. The constant part originates from a non-periodic component of vector potential $\mathbf{A}_\text{np} = \frac{1}{2} \mu_0 \mathbf{M}_\text{s} \times \mathbf{r}$, while the periodic part of the magnetic field originates from $\mathbf{A}_\text{p}$ computed as a periodic solution of $\Delta \mathbf{A}_\mathrm{p}(\mathbf{r}) = -\mu_0 \mathbf{j}(\mathbf{r})$, where $\mathbf{j}(\mathbf{r})$ is the spin current density. By using a Gordon decomposition and neglecting orbital currents, the spin current density is \cite{Rother2009} $\mathbf{j}(\mathbf{r}) = \nabla \times \mathbf{m}(\mathbf{r})$, where the spin magnetization density $\mathbf{m}(\mathbf{r})$ is computed from electronic structure spin DFT calculations.

We expect that the microscopic variations of the magnetic field, $\mathbf{B}_\text{p}$, will only play a role in atomic resolution regime. For larger probes, such as EVBs with high OAM, effects of these variations average to zero and only the constant part of the magnetic field $\mathbf{B}_\text{c}$ will influence the scattering on top of the Coulomb potential. The situation can then essentially be understood in terms of Eq.~\ref{PeqConsB}.
 
\begin{figure}[hbt!]
	\centering
	\includegraphics[width=0.23\textwidth]{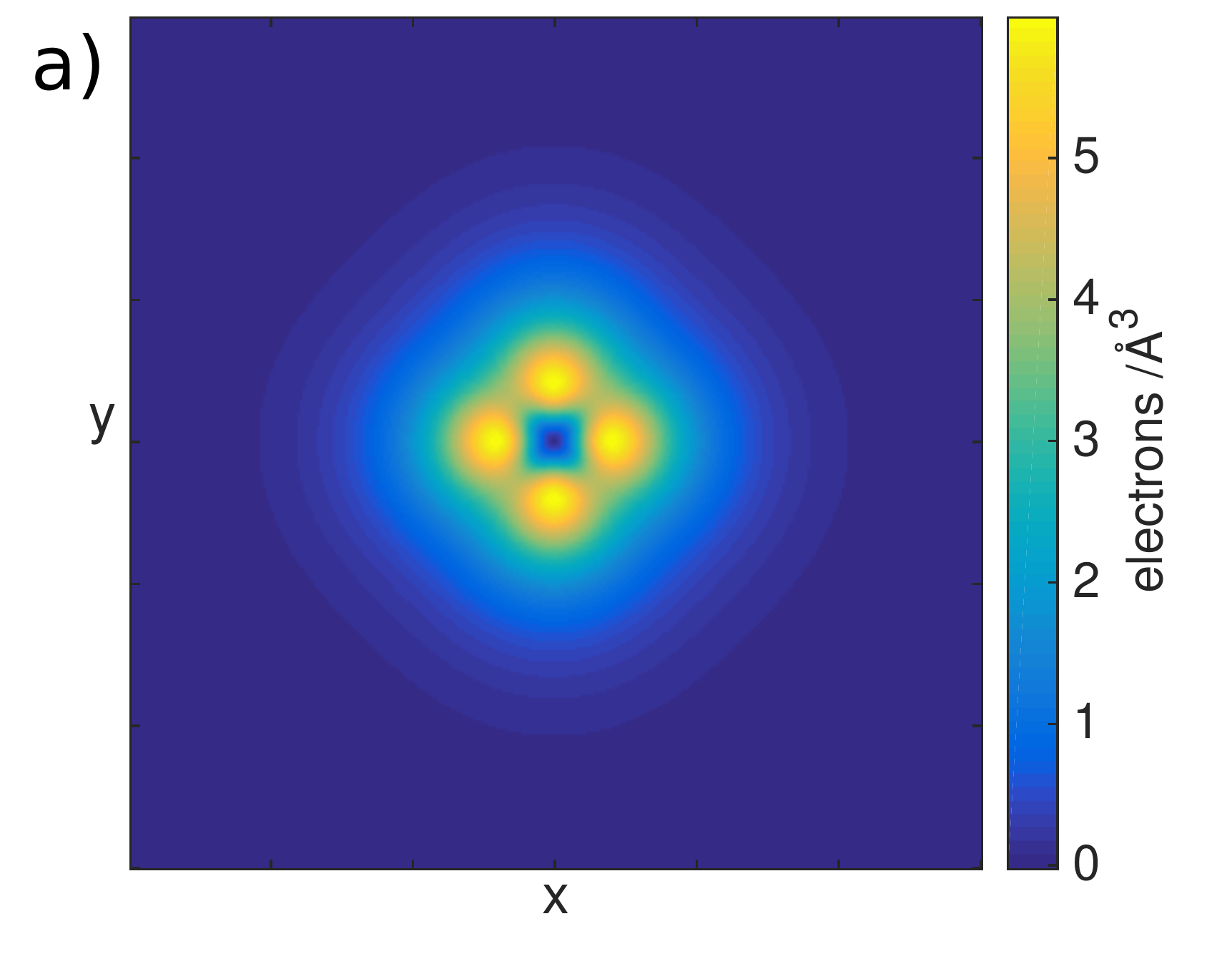}
	\includegraphics[width=0.23\textwidth]{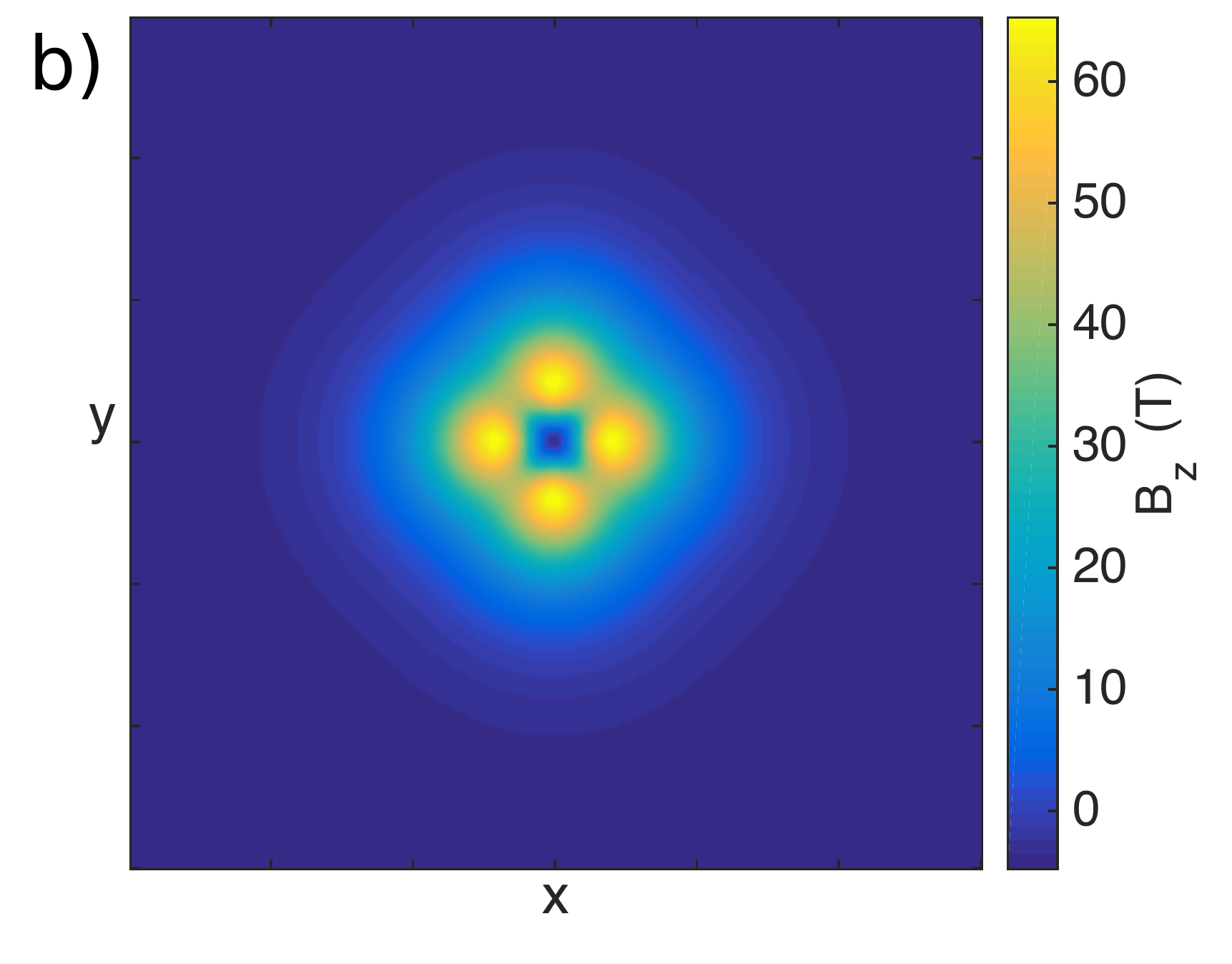} \\ 
	\includegraphics[width=0.23\textwidth]{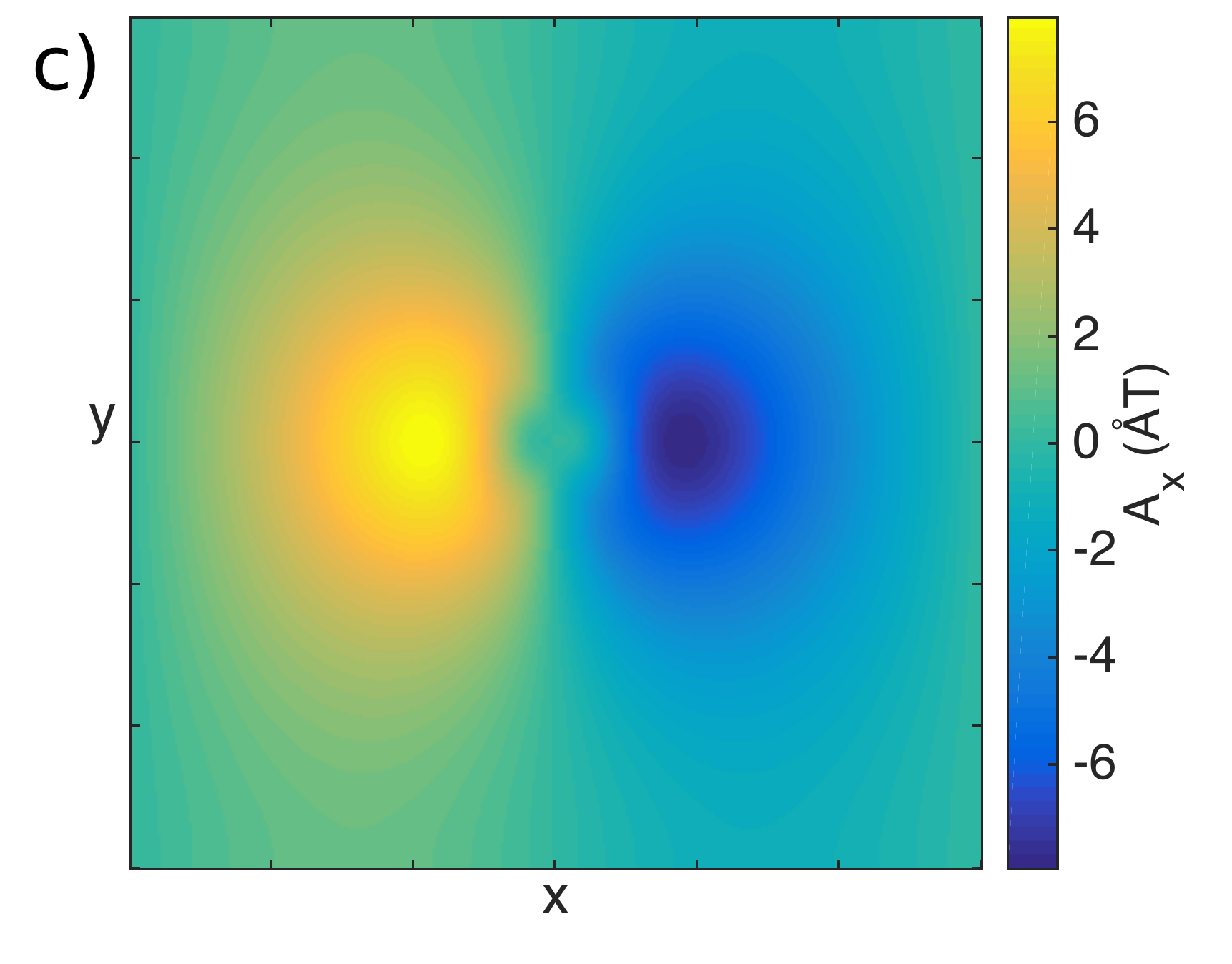}
	\includegraphics[width=0.23\textwidth]{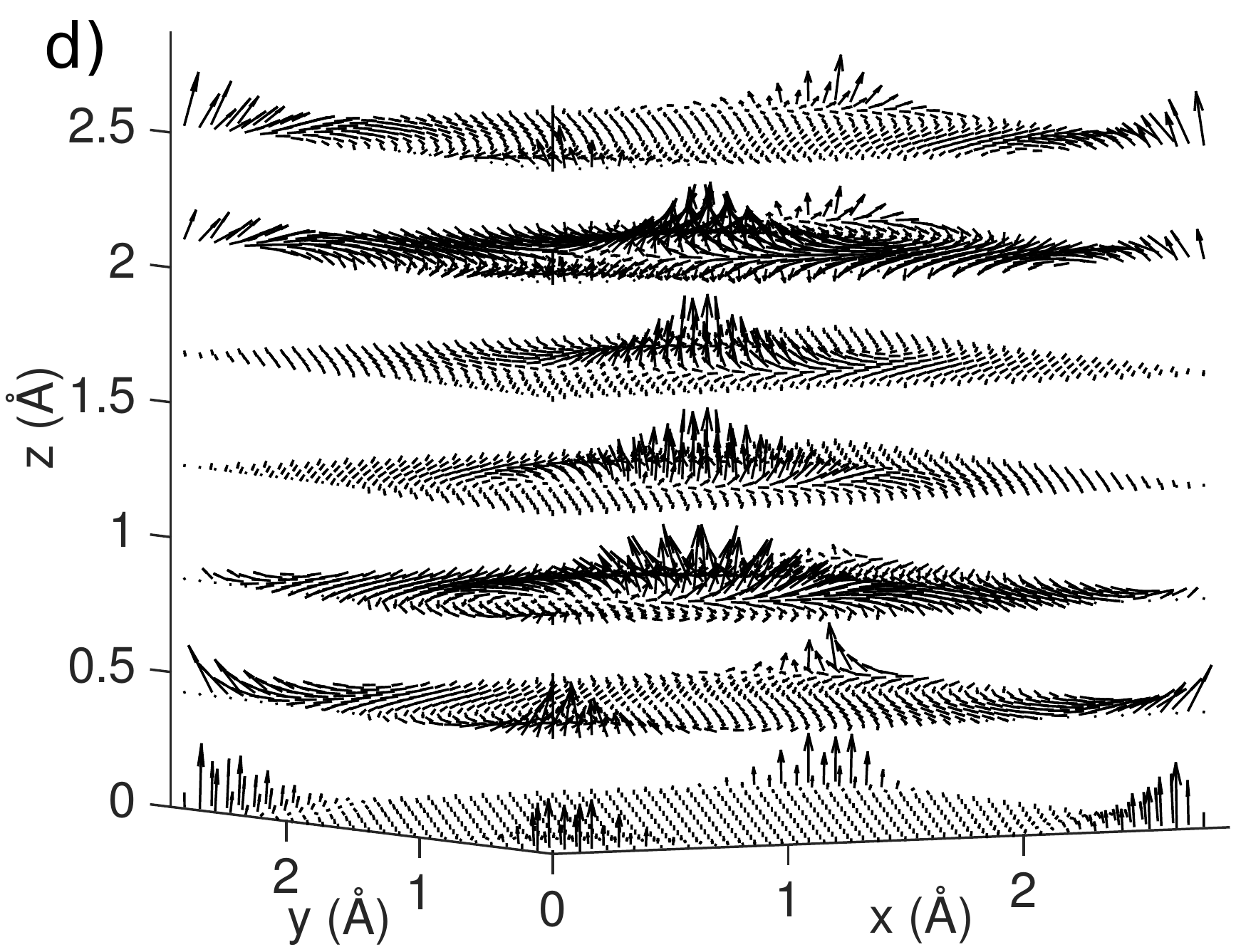} \\ 
	\caption{a) Spin magnetization density, b) $z$-component of the $\mathbf{B}$-field, c) $x$-component of the $\mathbf{A}_\text{p}$-field and d) the $\mathbf{B}$-field in a unit cell of bcc Fe as obtained from DFT calculations.}
	\label{fig.fields}
\end{figure}

Results of the procedure described above, applied to bcc Fe, are illustrated in Fig.~\ref{fig.fields}. In the case of collinear magnetism, $\mathbf{m}(\mathbf{r})$ is parallel to the $z$-direction, whereby $\mathbf{j}(\mathbf{r})$, and in the gauge chosen here also $\mathbf{A}(\mathbf{r})$, have non-zero $x$- and $y$-components only. The spin magnetization density as obtained via a collinearly spin-polarized full-potential linearized augmented plane wave \cite{Blaha2001} calculation in the generalized gradient approximation \cite{Perdew1996} is shown on an $xy$-cross section containing the central Fe atom of the bcc unit cell, Fig.~\ref{fig.fields}a. The $z$-component of the $\mathbf{B}$-field, Fig.~\ref{fig.fields}b, and the $x$-component of the $\mathbf{A}_\mathrm{p}$-field, Fig.~\ref{fig.fields}c, are plotted within the same plane. Note that the $z$-component of $\mathbf{B}_\text{p}$ reaches values in the order of 60~T---significantly larger than $\mu_0 M_\text{s}=2.2~\text{T}$ in bcc Fe. $A_y$ (not shown) is identical to $A_x$ rotated by $90^{\circ}$ about the $z$-axis and $A_z=0$ everywhere. Finally, the microscopic $\mathbf{B}_\text{p}$-field (to which a $B_\text{c}= 2.2~\text{T}$ in the $z$-direction should be added) is plotted as a vector field in one unit cell, Fig.~\ref{fig.fields}d. Although the shape of the spin density is very similar to that of $B_z$, they are not identical, and, even though only collinear spin density along the $z$-direction is considered, the $\mathbf{B}_\text{p}$-field has non-zero $x$- and $y$-components.

Due to persisting limitations in the creation of EVBs, electron beams with large OAMs ($\gg 1\hbar$) cannot be focussed on an atomic scale. We thus concentrate first on a situation, where we do not aim for atomic resolution, but rather enhance the Zeeman interaction by a large initial OAM ($=l\hbar$) of the beam. Using Eq.~\ref{multislicesol} we propagate electron beams with an initial OAM of $20\hbar$, $30\hbar$ and $40\hbar$, through a bcc Fe crystal of thickness up to 400 unit cells (115~nm). The radial shape of the beams is described by
\begin{equation}
\psi_l (k_\perp, \phi) \sim \ee^{\img l \phi} \Theta(\alpha k - k_\perp), 
\end{equation}
where $k_\perp$ and $\phi$ are cylindrical coordinates in $\mathbf{k}$-space and $\alpha$ is the convergence semi-angle. The lateral supercell dimension was $48 \times 48$ unit cells and each unit cell was discretized on a $64 \times 64 \times 64$ pixel grid. The acceleration voltage was $200~\text{kV}$ and $\alpha = 10$~mrad, corresponding to outer full-widths at half-maximum of $2.2$, $3.1$, and $3.9$~nm for $l=20$, 30, and 40, respectively. Beams were centered on an atomic column, but we have verified that the results do not depend on the exact beam position, as is expected for beams with spatial extent significantly larger than the crystal unit cell.

\begin{figure}[hbt!]
	\centering
	\includegraphics[width=0.5\textwidth]{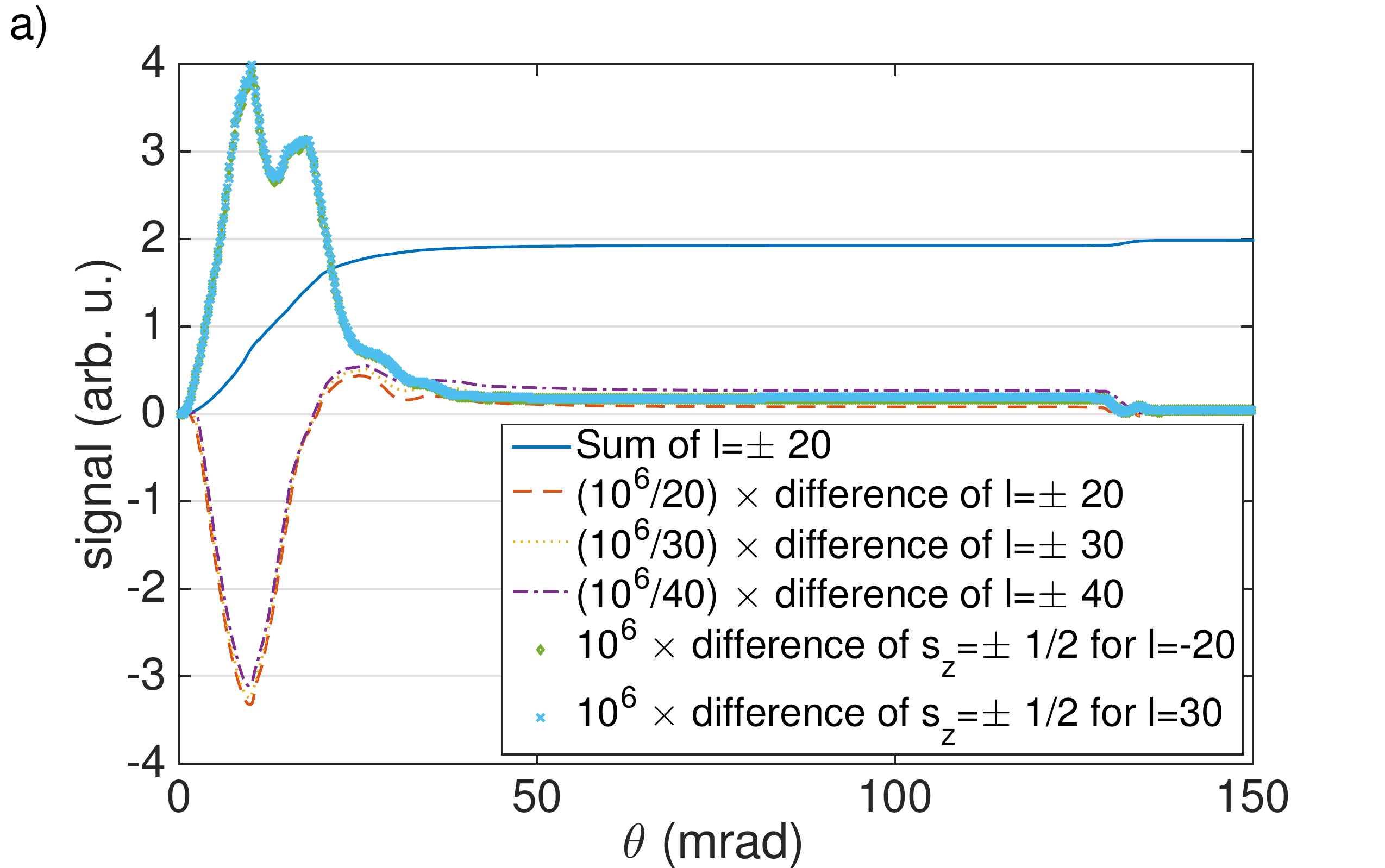} \\ 
	\includegraphics[width=0.5\textwidth]{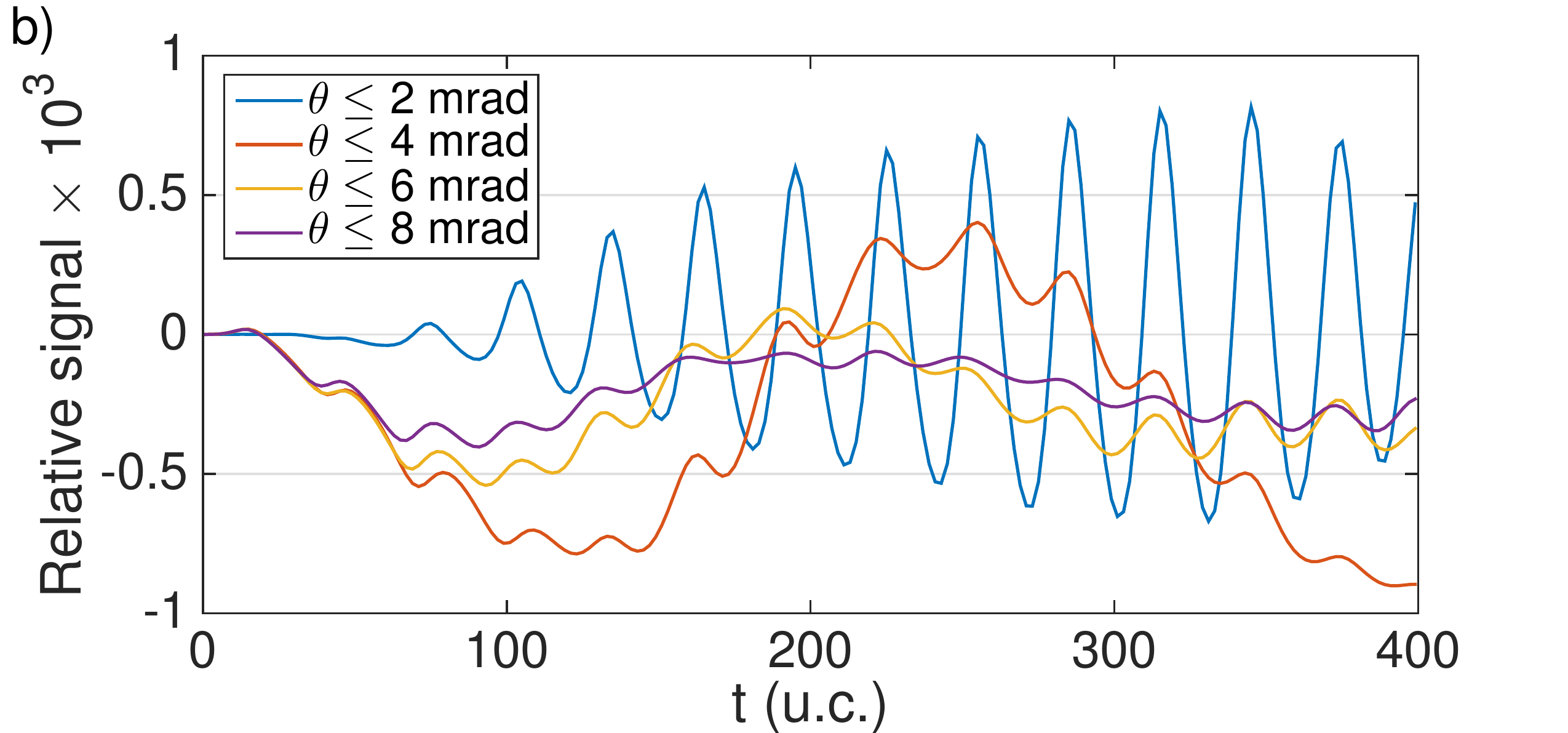}
	\caption{
	a) Magnetic signal (difference of intensities) as a function of collection angle. Differences are shown for unpolarized beams with $l=\pm20, \pm30$ and $\pm40$, as well as beams with $l$ fixed to $-20$ or $30$, but with difference taken over opposite spin channels. For $l=\pm 20$ the total signal (sum of intensities for opposite OAM) is also shown. Sample thickness was set to $t=42$ unit cells (12~nm). b) Thickness dependence of the relative magnetic signal for a circular collection aperture with collection angle indicated ($l=\pm 30$).
	}
	\label{fig.largeOAM}
\end{figure}  

A non-spin-polarized beam is in a mixed state, with 50\% of electrons with spin-up and 50\% spin-down, respectively. Therefore each simulation consists of two runs, one for each spin orientation, and the resulting diffraction patterns were averaged over the two spin orientations. It is worth mentioning that the proportion of spin-up electrons scattering into spin-down states or vice versa is negligible (of the order $10^{-14}$), although it has been suggested that for magnetization in the $xy$-plane the spin-flip scattering can be more significant \cite{GrilloKarimi}.

The Zeeman interaction leads to a redistribution of intensity in the diffraction pattern. Total intensity of scattered electrons is of course the same for positive or negative OAM, but the intensity of electrons scattered to smaller or larger angles varies depending on the OAM (see Fig.~\ref{fig.largeOAM}a). For large collection angles the intensity of scattered electrons saturates at a value of one, as dictated by normalization of the initial probe wavefunction. The intensity difference for, e.g., $l=\pm 20$ shows a peak around a collection angle of 10~mrad, after which its amplitude decreases eventually reaching zero. Computing such differences in a simulation with zero magnetic fields merely yields a numerical noise around ten orders of magnitude smaller. The kink observed close to $\theta = 130~\text{mrad}$ appears near a higher order Laue zone \cite{Juchtmans2015}.

Notice, how the magnetic signal is approximately proportional to the initial OAM of the EVB. This suggests that the strength of this signal can be further scaled up for beams with larger OAM. Consistently with Eq.~\ref{PeqConsB}, the magnetic signal obtained as an intensity difference from opposite spin channels, but for a fixed initial OAM, is 1) independent of OAM, and 2) of the same order of magnitude as the intensity difference due to changing the sign of OAM, when normalized per unit of OAM.

In Fig.~\ref{fig.largeOAM}b) the magnetic signals, for a disc shaped region with collection angles indicated in the legend, are shown for the $l=\pm 30$ case as a function of sample thickness. Accordingly, the magnetic signal can become significantly stronger for thicker samples. After 100 u.c. significant values of the relative magnetic signal in the range of $10^{-3}$ are observed. Considering that the signal is proportional to OAM and OAMs of size several hundreds have been reported \cite{Grillo2015}, signal strengths of few percent can be reached. While the intensity differences sensitively depend on the shape of the probe, magnetism should be measurable in an experimental setup with symmetrical $+l$ and $-l$ OAM probes --- such as those generated by holographic zone plates \cite{McMorran2011, Verbeeck2010}. Relative signals of several percent are well within the the detection limits of current bright field EVB STEM experiments, which easily integrate order of $10^6$ electrons (for typical probe currents, dwell times, collection angles) with a corresponding shot noise order of $10^3$, i.e., $1$\textperthousand.

Now we turn our attention to the atomic resolution regime. As was mentioned above, to focus EVB with large OAM onto atomic scales requires very large convergence angles in the range of hundreds of milliradians, which is outside of present instrumental possibilities. Therefore, we restrict ourselves to EVBs of small initial OAM, thereby reducing the magnitude of the Zeeman term. However, we remind that the strength of local microscopic magnetic fields can reach substantial values of several tens of Teslas (see Fig.~\ref{fig.fields}b), which could potentially lead to significant magnetic signals even in the atomic resolution regime. To assess the effect, we have performed simulations using beams with $l=\pm1$ and a rather large convergence angle of $40~\text{mrad}$ at an acceleration voltage of 300~kV. The supercell dimension was $24\times24\times100$ unit cells of bcc Fe (28.7~nm thick), each cell discretized on a $112\times112\times112$ grid. The collection angle was set to 5~mrad.

\begin{figure}[hbt!]
	\centering
	\includegraphics[width=0.5\textwidth]{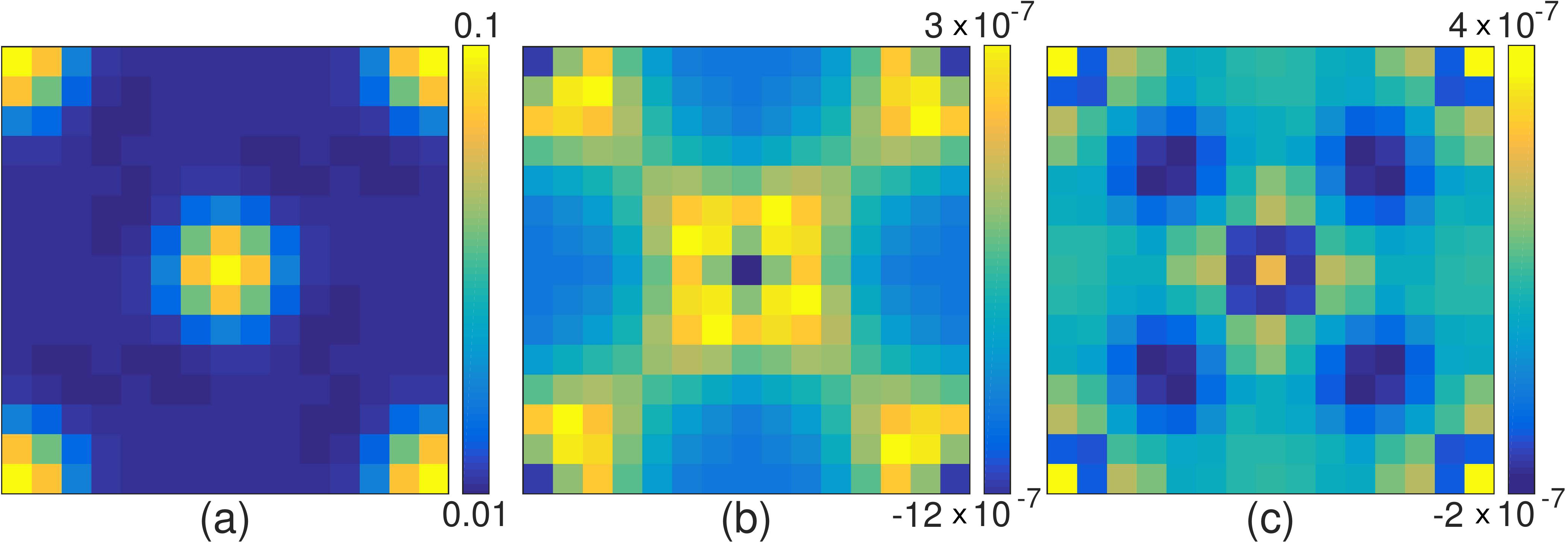}
	\caption{Simulated STEM image for a collection angle 5~mrad, as obtained from a) the spin averaged beam for $l=+1$, b) magnetic signal obtained as difference signal due to OAM $l = \pm 1$, c) difference between spin-up and spin-down beams for $l=+1$.}
	\label{fig.atoms}
\end{figure} 

Fig.~\ref{fig.atoms} summarizes atomic resolution simulations. The STEM image (Fig.~\ref{fig.atoms}a) should be compared to the magnetic signal computed from the $l=\pm 1$ difference (Fig.~\ref{fig.atoms}b) and spin-difference at $l=1$ (Fig.~\ref{fig.atoms}c). Note that a mirror image of vortex beam with OAM equal to $l\hbar$ is a vortex beam with OAM equal to $-l\hbar$ \cite{Rusz2014}. For this reason the magnetic signal is obtained as a difference of intensity of $l=+1$ beam at position $(x,y)$ and $l=-1$ beam at $(y,x)$.

Both spin and OAM differences are of the same order of magnitude, which is about $10^{-5}$ of the total signal. In the following we explore some routes to enhance the magnetic signal. An increase in its relative strength can be achieved by a further optimization of sample thickness, convergence angle, acceleration voltage, initial OAM and collection angle. To find the global maximum of this multidimensional optimization problem is a formidable task, which will furthermore depend on the material to be investigated. Therefore, we focus on a few parameters only and assess the increase in magnetic signal in an approximative manner.

\begin{figure}[hbt!]
	\centering
	\includegraphics[width=0.5\textwidth]{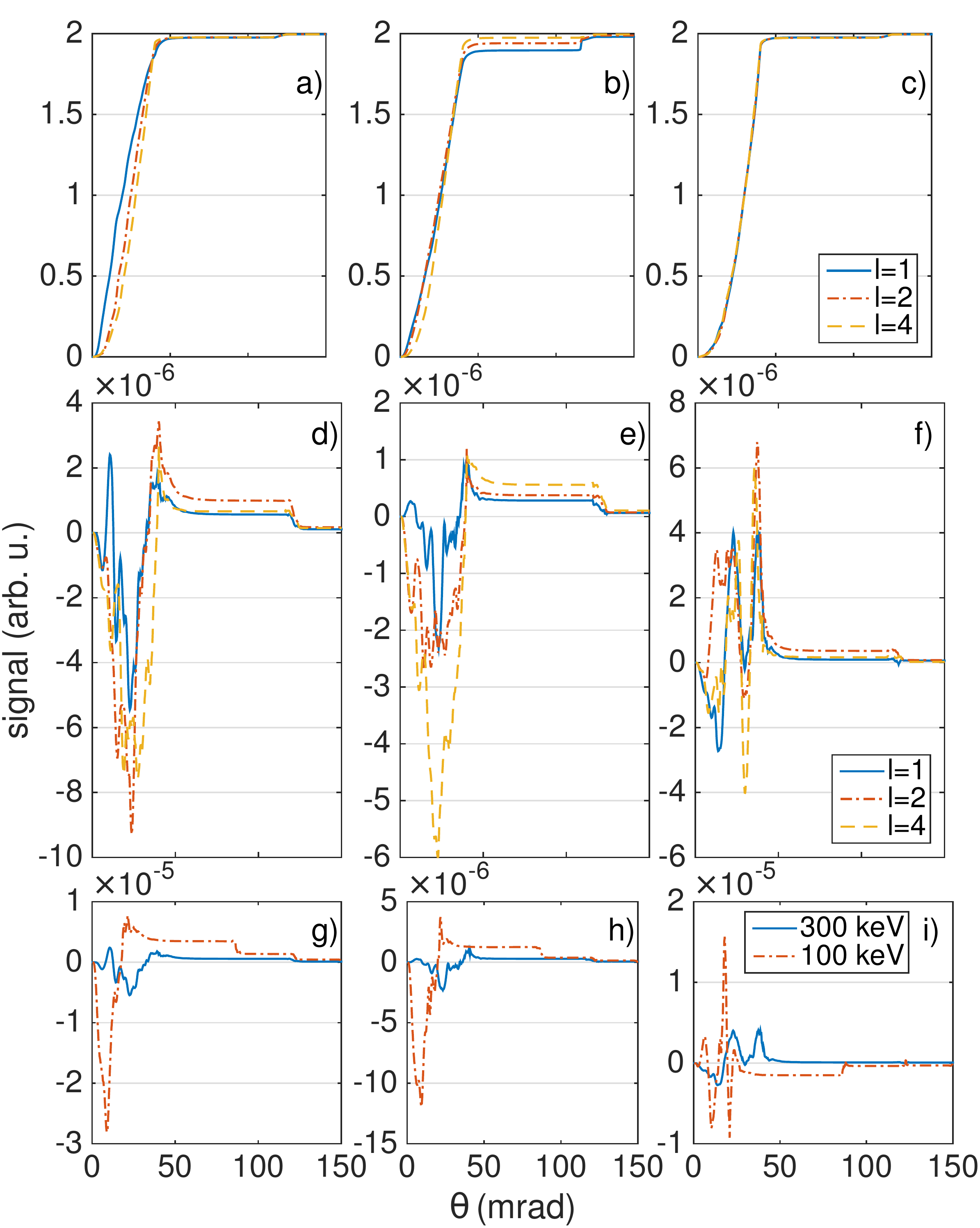} \\ 
	\caption{Sum in a)-c) and difference in d)-f) of signal integrated over collection angles from $0$ to $\theta$ for $+l$ and $-l$ beams with $l=1$, $2$ and $4$ at beam positions $(0,0)$ in a), d), g), $(1,1)\frac{a}{14}$ in b), e), h) and $(1,0)\frac{a}{2}$ in c), f), i). g)-h) show the difference between acceleration voltage of 100 keV or 300 keV with other parameters kept same for the three respective beam positions.}
	\label{fig.smallOAM}
\end{figure}

To investigate the effect of larger OAM on the magnetic signal strength, a few beam positions were recalculated with with $l=\pm 2$ as well as $l=\pm 4$, see Fig.~\ref{fig.smallOAM}a)-f). Accordingly, the proportionality between magnetic signal and $l$ is lost in the atomic resolution regime. This is most likely due to a different form of the magnetic interaction $\propto \mathbf{A}\cdot\mathbf{p}$, which is not anymore directly proportional to the OAM and depends on details of spatial distribution of the magnetic field and the probe wavefunction. Note for example the radial intensity profiles for beams with $l=1,2,4$. The differences are mostly due to strong pinning of beams with low OAM to atomic columns \cite{Lubk2013}, less pronounced for OAM=$2\hbar$ and $4\hbar$, respectively. Nevertheless, magnetic signals are still somewhat stronger for beams with larger OAM.

An alternative route to enhancing the magnetic signal consists of reducing the acceleration voltage. Upon inspection of Eq.~\ref{paraxialPeq} we note an additional prefactor $\gamma^{-1}$ in front of the magnetic coupling compared to the electric one resulting in a relative increase of the magnetic signal at lower acceleration voltages. Fig.~\ref{fig.smallOAM}g)-i) compares results obtained with voltages 100~kV and 300~kV for beams with initial OAM of $\pm \hbar$ and other parameters kept fixed. An increase of magnetic signal by a factor of $\sim 3$ or 4 can be observed for the lower acceleration voltage. 

Combining all the effects a further optimization of all parameters is suited to increase the relative magnetic signal strength by one order of magnitude to $10^{-4}$. Yet, the relative magnetic signal strength of $10^{-4}$ means that it will be extremely sensitive to scan noise, drifts and changes of sample orientation during data acquisition, which renders atomic resolution measurements of a magnetic signal based on the Zeeman interaction of OAM with magnetic fields in the sample extremely challenging, most likely beyond the possibilities of present instruments.

In conclusion, we have demonstrated computationally that the elastic scattering of electron vortex beams on magnetic samples in the TEM depends on the relative orientation of the initial OAM and the magnetization in the sample. In principle, this effect opens a new way for measurement of magnetic properties. For beams with OAM of few hundreds of $\hbar$, the predicted relative strength of magnetic signal should reach up to a few percent, calling for an experimental verification. If successful, this permits a new way of characterization of magnetic properties at about 10~nm spatial resolution. In the atomic resolution regime, the calculated relative magnetic signal strength reaches only up to $10^{-4}$, making it unlikely to be detected with present-date instruments, particularly due to scan noise and unavoidable sample drifts.

\acknowledgments
AE and JR acknowledge Swedish Research Council and G\"oran Gustafsson's Foundation for financial support. AL acknowledges financial support from the European Union under the Seventh Framework Program under a contract for an Integrated Infrastructure Initiative (Reference 312483 - ESTEEM2). Valuable discussions with Nobuo Tanaka, Jo Verbeeck and Vincenzo Grillo are gratefully acknowledged.

\bibliography{literature}{}

\begin{thebibliography}{35}
\bibitem[{Asq()}]{Asquared}
\bibinfo{note}{A term proportional to $A^2$ is neglected as a higher order
  relativistic correction.}
\end{thebibliography}
\bibliographystyle{apsrev}

\end{document}